\documentstyle[12pt]{article}
\begin{document}

\centerline{\bf ON THE VACUUM ENERGY DENSITY}
\centerline{\bf AND NON-PERTURBATIVE EFFECTS IN SIGMA MODELS}

\vspace{10mm}
\centerline{\bf V.G.Ksenzov}

\vspace{10mm}
\centerline{Institute of Theoretical and Experimental Physics, Moscow}

\vspace{20mm}
\begin{abstract}
The vacuum energy density is obtained for the $O(N)$ nonlinear sigma model.
It is shown that non-perturbative contributions are connected with the square
of the symmetry current of the group $O(N)$. This result is valid for
$\sigma$-
fields which are subject to the constraint.
\end{abstract}

After the discovery of nonperturbative fluctuations the question of
their role in generating the physical amplitude has been of particular
interest for the field theory. A lot of effort has been invested in
investigating these effects in Yang-Mills theory, but the question is still
far from being solved. Therefore it makes sense to study these problems
in a simple theory in order to get a new insight into the subject.
Such a simple analogy to the Yang-Mills theory is the two-dimensional sigma
model with fields transforming according to the vector representation of the
group $O(N)$ [1]. This model provides a possibility to perform an analysis,
and to get the results which might be relevant to actual physics.

For a long time it has been clear that the dynamics is in some way
related to a complicated vacuum structure, therefore investigation of the
vacuum structure turns out to be the key to the understanding of the actual
dynamics.

In this paper we will discuss the influence of non-perturbative
fluctuations on the vacuum energy density $\varepsilon_{vac}$ in the
framework of the two-dimensional sigma model in the large $N$ limit.
In these models non-perturbative effects occur which do not depend
on the existence of instantons for $N>3$.

We first recall what is known
about nonperturbative effects and $\varepsilon_{vac}$ in sigma model.
These results were first obtained in ref.[2].
Lorentz invariance implies that the vacuum energy density may be defined as
$$g_{\mu\nu}\varepsilon_{vac}= <0\mid \theta_{\mu\nu}\mid 0> ~,$$
where $\theta_{\mu\nu}$ stands for the energy-momentum tensor. Taking
into account the conformal anomaly we arrive at
$$\varepsilon_{vac}=\frac{1}{2} <0\mid \theta_{\mu\mu}\mid 0> ~.$$

In the large $N$ sigma model the regularized expression for
$\theta_{\mu\mu}$
may be written as a sum of two terms
$$\varepsilon_{vac}=\frac{N}{8\pi} (-M^2+\frac{<\alpha>}{\sqrt{N}}) ,$$
where $M^2$ is the regulator mass, $\alpha$ is the Lagrange multiplier
and $<\alpha>$ is a constant denoted elsewhere by $m^2 \sqrt{N}$.
The first term is connected
with the perturbative fluctuations in the vacuum, while the second one
is due to non-perturbative fluctuations. Let us note that $<\alpha>$ is
a fundamental parameter of the
theory also in the case of phase transitions [3].

To demonstrate how the non-perturbative effects enter into the value of
$<\alpha>\not= 0$ one employs  the Lagrangian
\begin{equation}
L=\frac{1}{2} \{(\partial _{\mu}\sigma^a)^2 + \frac{\alpha}{\sqrt{N}}
(\sigma^a \sigma^a - \frac{N}{f})\}~~.
\end{equation}
We want to decompose $\sigma^a(x)$ and $\alpha(x)$ as
\begin{eqnarray}
\sigma^a(x) = c^a(x) + q^a(x)\\ \nonumber
\alpha(x) = \alpha_c(x) + \alpha_{qu}(x)~~,
\end{eqnarray}
where $c^a(x)$ and $\alpha_c(x)$ are classical fields, while
$q^a(x)$ and $\alpha_{qu}(x)$ describe small fluctuations around the
classical background.

The equations of motion for $c^a$ and $\alpha_c$ fields are
\begin{equation}
\partial ^2 c^a= \frac{\alpha c^{a}}{\sqrt{N}}
\end{equation}
\begin{equation}
c^a(x) c^a(x) = \frac{N}{f}
\end{equation}
\begin{equation}
\frac{\alpha_c(x)\sqrt{N}}{f} = -(\partial_{\mu}c^a)^2 ~.
\end{equation}

The equation (5) may be rewritten in a linear form in $(\partial_{\mu}c^a)$.
Using the symmetry current $J_{\mu}^{ab}$ and the constraint (4) we get
\begin{equation}
\frac{1}{2}(J_{\mu}^{ab})^2 = \frac{N}{f}(\partial_{\mu}c^a)^2 ~,
\end{equation}
where $J^{ab}_{\mu}= c^a \partial_{\mu} c^b - \partial_{\mu} c^a~ c^b$~, and

\begin{equation}
-\frac{\alpha_c(x)}{\sqrt{N}}= \frac{1}{2}(\frac{f}{N} J_{\mu}^{ab})^2
\end{equation}
The Lagrange multiplier turns out to be connected with the square of the
symmetry current of the group $O(N)$.

By virtue of eqs.(5) and (7) we can obtain a new nonlinear equation
$$\partial_{\mu} c^a (\partial_{\mu} c^a + \frac{f}{N} J^{ab}_{\mu} c^b)
= 0$$~.

Then we can obtain two simple equations
$$ \partial_{\mu} c^a = 0$$
or
\begin{equation}
\partial_{\mu} c^a + \frac{f}{N} J^{ab}_{\mu} c^b = 0~.
\end{equation}
The solutions of these equations are special forms of the solutions
of equation (5).
The solution of the first equation can not give non-perturbative effect
because the
Lagrange multiplier is zero (5). The eq.(8) is identically satisfied
provided
$c^a (x)$ are subject to the constraint (4).
These fields have a special property: their transformation
under translations can be compensated by isotopic
transformations as it was first discussed in [4].

The classical energy-momentum tensor can be found from (8) and we arrive at
\begin{equation}
\theta_{\mu\nu}=\frac{f}{N}(J_{\mu}^{a}J_{\nu}^a -
\frac {g_{\mu\nu}}{2}(J_{\mu}^a)^2)
\end{equation}
with the evident properties
$$\partial_{\mu}\theta_{\mu\nu}= 0~, ~~ \theta_{\mu\mu}= 0~.$$

The energy-momentum tensor $\theta_{\mu\nu}$ turns out to be proportional
to the product of conserved currents in this model. Similar facts are
well known in field theory [5] and in quantum mechanics [6].

To find the conformal anomaly $\theta_{\mu\mu}$ we define it as
\begin{equation}
\theta_{\mu\mu} = lim <0\mid M^2 q^a(x-\Delta)q^a (x+\Delta) \mid 0> ~.
\end{equation}
where $M^2$ is the regulator's mass and $q^a$ is the regulator's field.
The limit of the right-hand-side of Eq. (10) is calculated at
$M^2 \rightarrow \infty$ and $\Delta^2 \rightarrow 0$ and under a
special condition $\Delta^2 M^2 = Const$.

In equation (10) fast varying  fluctuations $q^a(x)$ are not  arbitrary
but must satisfy the constraint [2]
\begin{equation}
2c^a q^a + q^a q^a = 0 ~.
\end{equation}

This contraint can be used in linear form in $q^a$ at large $N$ [2].
Differentiating $c^a q^a = 0$ and using (8) we get
\begin{equation}
\partial_{\mu}q^a = - \frac{f}{N} J_{\mu}^{ab}q^b
\end{equation}

Equation (12) connects quantum field fluctuations and non-perturbative
fluctuations.

Our goal is to calculate (10) and we can do this by knowing
equation (12) alone. Indeed, we can obtain
$$\theta_{\mu\mu} = lim \frac{1}{2} (<0\mid M^2 q^a q^a\mid 0> - $$
\begin{equation}
- \Delta_{\mu} \Delta_{\nu} M^2
 <0\mid q^a q^k\mid 0> \left( \frac{f}{N} \right)^2 J_{\mu}^{ab}
J^{bk}_{\nu} + \ldots)
\end{equation}
The first term is the perturbative contribution to $\theta_{\mu\mu}$.
Knowing the value of this term we can conclude that
$$< 0 \mid q^a q^k \mid 0 > = - \frac{\delta^{ak}}{4\pi}$$~.
The vacuum expectation value of $\theta_{\mu\mu}$ or the vacuum energy
emerges
after identifying the function of the classical fields with the vacuum
expectation value of the corresponding operator built from $\sigma^a$.
Then we get
$$ <0\mid(\frac{f}{N})^2 J_{\mu}^{ab}J^{bk}_{\nu}\mid 0>= $$
$$ = -g_{\mu\nu}\delta^{ak}
<0\mid (\frac{f}{N})^2 J_{\lambda}^{2}\mid 0>= g_{\mu\nu}\delta^{ak}
< 0 \mid \frac{\alpha (x)}{\sqrt{N}} \mid 0 > = m^2 g_{\mu\nu}
\delta^{ak} $$
We have used eq.(7) and the stationary value of $\alpha(x)$.

In this paper, we have concentrated on the non-perturbative contribution
to the vacuum energy density. It was shown that non-perturbative
fluctuations
whose transformation under translations can be compensated by
isotopic transformations
determine non-perturbative contribution into $\varepsilon_{vac}$.
As a matter of fact, vacuum condensate of $<\alpha>$ is vacuum condensate of
the square of the symmetry current of the group $O(N)$ buit from $\sigma$-
fields which are in turn subject to the second class constraint.

The author is grateful to V.M.Belyaev and V.A.Novikov
for useful discussions.

\end{document}